\documentclass[aps,preprint,preprintnumbers,nofootinbib,showpacs,prd]{revtex4-1}
\usepackage{graphicx,color,amsmath,amsfonts}

\newcommand{\beq}     {\begin{equation}}
\newcommand{\eeq}     {\end{equation}}
\newcommand{\bea}     {\begin{eqnarray}}
\newcommand{\eea}     {\end{eqnarray}}
\newcommand{\pmns}{\mbox{$ U_{\rm PMNS}$}}
\newcommand{\bad}{\begin{array}{ccc}}
\newcommand{\ea}{\end{array}}

\newcommand{\gauge}{$SU(3)_C \!\times\! SU(2) _L\!\times\! SU(2)_R\!\times\! U(1)_{X}$}
\newcommand{\lsim}{\mathrel{\mathop{\kern 0pt \rlap
  {\raise.2ex\hbox{$<$}}}
  \lower.9ex\hbox{\kern-.190em $\sim$}}}
\newcommand{\gsim}{\mathrel{\mathop{\kern 0pt \rlap
  {\raise.2ex\hbox{$>$}}}
  \lower.9ex\hbox{\kern-.190em $\sim$}}}
  
\newcommand{\no}     {\nonumber}
\newcommand{\di}      { \mathrm{d }}

\newcommand{\lm}      {\lambda}
\newcommand{\es}      {\epsilon}

\newcommand{\dt}      {\delta}
\newcommand{\Dt}      {\Delta}
\newcommand{\kp}      {\kappa}

\newcommand{\gev}      {\;{\rm GeV}}
\newcommand{\tev}      {\;{\rm TeV}}

\def\n{{(n)}}

\def\fan{{f_A^{(n)}}}
\newcommand{\lmt}{{\lambda_t}}
\newcommand{\acp}    {A_{ CP}}
\begin{document}

\title{The $B \to \pi K $ puzzle and the Bulk Randall-Sundurm model
}

\author{Sanghyeon Chang}
\email{schang@konkuk.ac.kr}
\affiliation{%
School of Physics, Konkuk University,
                   Seoul 143-701, Korea
}
\author{C.~S. Kim}
\email{cskim@yonsei.ac.kr, Corresponding Author}
\affiliation{
Department of Physics and IPAP, Yonsei University,
Seoul 120-749, Korea}
\author{Jeonghyeon Song}%
 \email{jhsong@konkuk.ac.kr}
\affiliation{%
Division of Quantum Phases \& Devices, School of Physics,
Konkuk university, Seoul 143-701, Korea
}
\date{\today}

\begin{abstract}
\noindent The recent measurements of the direct CP asymmetries
($A_{CP}$) in the
penguin-dominated $B\to K \pi$ decays
show some discrepancy from the standard model (SM)
prediction.
While $A_{CP}$ of $B^+ \to
\pi^0 K^+$ and that of $B^0 \to \pi^- K^+$
in the naive estimate of the SM
are expected to have very similar values,
their experimental data are of the opposite sign and
different magnitudes.
We study the effects of the custodial bulk Randall-Sundrum model
on this $A_{CP}$.
In this model, the misalignment of
the five-dimensional (5D) Yukawa interactions
to the 5D bulk gauge interactions in flavor space
leads to tree-level
flavor-changing neutral current by the Kaluza-Klein
gauge bosons.
In a large portion of the parameter space of this model,
the observed nonzero
$A_{CP}(B^+ \to
\pi^0 K^+) - A_{CP}(B^0 \to \pi^- K^+)$
can be explained 
only with low Kaluza-Klein mass scale $M_{KK}$
around $1$ TeV.
Rather extreme parameters is required to explain it
with $M_{KK}\simeq 3\tev$.
The new contributions to well-measured branching ratios of $B \to K \pi$ decays
are also shown to be suppressed.

\end{abstract}

\maketitle

\section{Introduction}
\label{sec:introduction}

The study of $B$ meson decays at Belle and BaBar~\cite{exp}
have been a crucial probe of the standard model (SM),
especially its CP violation part.
Recently the $B \to K \pi$ decay system has drawn a lot of
interest due to the discrepancy between the SM  predictions
and the measurements
\cite{Buras:2003yc,Gronau:1994rj,Kim:2007kx,Baek:2008vr,others}.
There are nine measurements of the four decays of
$B^+ \to \pi^+ K^0 $,
$B^+ \to \pi^0 K^+ $,
$B^0 \to \pi^- K^+ $,
and $B^0 \to \pi^0 K^0 $:
four branching ratios (BR),
four direct CP asymmetries $A_{\rm CP}$,
and one mixing-induced CP asymmetry $S_{\rm CP}$.
The 2008 data of these nine measurements
are in Table \ref{tab:b2pik}.

\begin{table}[tbh]
\center
\begin{tabular}{|c|ccc|}
\hline
Mode & $BR\;[10^{-6}]$ & $A_{\rm CP}$ & $S_{\rm CP}$ \\ \hline
~~~$B^+ \to \pi^+ K^0$~~~\ \ & \ \ ~~~$23.1 \pm 1.0$~~~ &\ \ ~~~$0.009 \pm 0.025$~~~ & \\
$B^+ \to \pi^0 K^+$ & \ \ $12.9 \pm 0.6$ &\ \ $0.050 \pm 0.025$ & \\
$B^0 \to \pi^- K^+$ & \ \ $19.4 \pm 0.6$ &\ \ $-0.098^{+ 0.012}_{-0.011}$ & \\
$B^0 \to \pi^0 K^0$ & \ \ $9.8 \pm 0.6$ &\ \ $-0.01 \pm 0.10$ &\ \
$0.57 \pm 0.17$ \\
\hline
\end{tabular}
\caption{Experimental data for $B\to \pi K$; BR's,  direct CP asymmetries $A_{\rm CP}$, and
mixing-induced CP asymmetry $S_{\rm CP}$~\cite{HFAG,piKrefs}.}
\label{tab:b2pik}
\end{table}

We focus on the direct CP asymmetries $A_{\rm CP}$
of $B^+ \to \pi^0 K^+$ and $B^0 \to \pi^- K^+$.
In the SM, the dominant contribution to each decay amplitude
comes from the
strong penguin contribution $P$.
The color-suppressed tree contribution $C$
may be smaller than the color-favored tree contribution $T$
by a factor of the small parameter $\lm = | V_{us}| \simeq 0.22$.
Therefore, both $B^+ \to \pi^0 K^+$ and $B^0 \to \pi^- K^+$ could
have $A_{\rm CP}$ given by the interference between
$T$ and $P$ in the leading order, as shown in Eq. (1).
The direct CP asymmetries of
two decay modes are expected to be the same size with the same sign within the naive estimate of the SM.
As can be seen in Table \ref{tab:b2pik},
however, the observation is quite different
from this naive SM prediction:
$\acp(B^+ \to \pi^0 K^+)$ is non-zero positive\footnote{
Albeit statistically less significant with larger error
compared with BELLE and BaBar data,
the CLEO collaboration observed negative mean value for
this asymmetry~\cite{HFAG}, $A_{\rm CP}(B^+ \to \pi^0 K^+)=-0.29 \pm 0.23 \pm 0.02$.}
while $\acp(B^0 \to \pi^- K^+)$ is negative.
Known as ``$B \to \pi K$ puzzle'',
this discrepancy
has brought extensive attentions, leading to
model-independent studies as well as
new physics (NP) effect studies in the literature.

In this paper,
we study this $\acp$ puzzle in the framework of
the custodial bulk Randall-Sundrum (RS) model\,\cite{Agashe:2003zs}.
In the simplest
implementation of the
RS model, featuring an $SU(2)_L \times U (1)_Y$ 
bulk gauge symmetry 
and a minimal brane-localized Higgs sector,
this $B \to K \pi$ puzzle was studied,
showing the difficulty to solve the puzzle 
under the experimental constraints \cite{Bauer:2009cf}.
As a five-dimensional (5D) warped model
with all the SM fields in the bulk
(except for the Higgs boson field),
the bulk RS model provides
very attractive explanations for both gauge
hierarchy and fermion mass
hierarchy\,\cite{Randall:1999ee,Goldberger:1999wh,Davoudiasl:1999tf,Chang:1999nh, Grossman:1999ra,
Gherghetta:2000qt,Huber:2003tu,Chang:2005ya,Agashe:2004ay,Agashe:2004cp,Kim:2002kk}.
To ensure the $SU(2)$ custodial symmetry,
we adopt the model with the bulk gauge symmetry of \gauge,
induced from AdS$_5$/CFT feature\,\cite{Agashe:2003zs}.

Since the 5D Yukawa interaction is not generally flavor-diagonal in this
model,
the flavor-changing-neutral-current (FCNC)
is generated at tree level,
mediated by Kaluza-Klein (KK) gauge
bosons\,\cite{FCNC:RS,Chang:2006si}.
The 5D Yukawa couplings $\lm^{f}_{5ij}$
and the bulk Dirac mass parameters $c_f$'s
determine all the FCNC processes in principle.
Without a priori information about the model parameters, however,
this model lacks the prediction power.

In our previous works \cite{Chang:2006si,Chang:2007uz}, we show that if we adopt two simple
and natural
assumptions, we can fix
the model parameters
and extract the necessary information for the calculations.
The first assumption is that the 5D Yukawa couplings
are universal,
$\lm_{5ij} \simeq \lm_5 \sim \mathcal{O}(1)$.
The second assumption is that
fine tuning is not allowed when explaining the observed SM mixing matrices (CKM and PMNS).
Here our restriction is at the level of
no order-changing by cancellation.
Mild fine tuning is permitted in this setup.
With the given $\lm_{5ij}$'s and $c_f$'s
based on the two assumptions,
we study the bulk RS model effects on the $B \to K \pi$ decay.
We will show that this model
can explain
the discrepancy between the observed $\acp$
in the $B \to \pi K$ decay and the SM prediction
with the KK mass scale around 1 TeV.
These NP effects give suppressed
contributions to the well observed BR's.

The organization of the paper is as follows.  In Sec. \ref{sec:B2piK}, we
briefly review the current status of the measurements
of four $B \to K \pi$ decays.
In Sec. \ref{sec:RS},
we summarize the custodial bulk RS model
and formulate the bulk fermion sector.
After presenting two naturalness assumptions,
we determine all the bulk Dirac mass parameters.
Section \ref{sec:effect} deals with the effects
of this model on $A_{\rm CP}$'s of $B\to K \pi$ decays.
We conclude in Sec. \ref{sec:conclusions}.

\section{Short review of $B \to \pi K$ Decays}
\label{sec:B2piK}

In the SM, the $B \to \pi K$ decays are dominated by the $\bar{b}\to\bar{s}$ QCD penguin
diagrams.
The electroweak penguin and the tree contributions are next dominant.
The current experimental data in Table~\ref{tab:b2pik}
show that the branching ratios are very precisely measured.
The observations are more precise than the SM
theoretical estimates such as QCD factorization
and the perturbative QCD\,\cite{pQCD}.

The $B \to \pi K$ decay amplitudes can be written
in terms of topological
amplitudes up to $\lambda^2$ scale:
\bea
A(B^+ \to \pi^+ K^0) &=& P^{'} -\frac{1}{3} P^{'C}_{\rm EW}
 , \label{eq:all2}\\
\sqrt{2} A(B^+ \to \pi^0 K^+) &=& -P^{'} - T' - P'_{\rm EW}
-\frac{2}{3} P^{'C}_{\rm EW} - C'  ,
\nonumber\\
A({B}^0 \to \pi^- K^+) &=& -P^{'}  -T'  -\frac{2}{3} P^{'C}_{\rm EW} , \nonumber\\
\sqrt{2} A({B}^0 \to \pi^0 {K}^0) &=&
P^{'} - P'_{\rm EW}  -\frac{1}{3} P^{'C}_{\rm EW}
-C'.
\no
\eea
The primes denote the $\bar{b}\to\bar{s}$ transition.
The color-favored (color-suppressed) tree
diagrams are represented by $T'$ ($C'$), and the $P^{'(C)}_{\rm EW}$ is the
electroweak (color-suppressed electroweak) penguin.

The penguin diagram $P'$ is the sum of three up-type ($u,c,t$)
quark contributions:
\bea
P'&=& \lambda_u \tilde{P}_u + \lambda_c \tilde{P}_c + \lambda_t \tilde{P}_t
\nonumber\\
&=& \lambda_t (\tilde{P}_t-\tilde{P}_c)+ \lambda_u (\tilde{P}_u-\tilde{P}_c)
\nonumber\\
&\equiv& P'_{tc} +P'_{uc},
\eea
where $\lm_i \equiv V^*_{ib}V_{is} (i=u,c,t)$,
and the unitarity of the CKM matrix is used for the
second equality.
Here, the phase of $\lambda_t\equiv V^*_{tb}V_{ts}$ is $\sim \pi$ within the SM.
We also expect the following hierarchies from
theoretical calculations in the SM~\cite{QCDF,PQCD_NLO,SCET}:
\bea
\begin{array}{cl}
O(1)   & \qquad |P'_{tc}|, \\
O({\lambda})  & \qquad |T'|, |P'_{\rm EW}|, \\
O({\lambda}^2)  & \qquad |C'|, |P'_{uc}|,|P^{'C}_{\rm EW}|.
\label{eq:hierarchy}
\end{array}
\eea
If we define
\bea
\label{eq:dACP}
\Dt A_{\rm CP} \equiv A_{\rm CP}(B^+ \to
\pi^0 K^+) - A_{\rm CP}(B^0 \to \pi^- K^+),
\eea
the SM predicts very small  $\Dt A_{\rm CP}$,
which is contradictory to the experimental data
in Table~\ref{tab:b2pik}.
This discrepancy possibly
suggests the existence of the NP contribution,
especially in the CP violating phases.
If the NP contribution is the source of the discrepancy,
it should be of the order of $\lambda$ or more.

The effective Hamiltonian for $B \to \pi K$ can be written in operator expansion\,~\cite{Buchalla:1995vs}:
\begin{eqnarray}
\label{eq:H:def:C}
{\cal{H}}_{eff} = \frac{G_F}{\sqrt{2}}\left(\sum_{p=u,c}
\lambda_p \left( C_1 Q_1^p+C_2 Q_2^p\right)
-\lambda_t \sum_{i=3}^{10} C_i Q_i \right).
\label{effective}
\end{eqnarray}
The operators are defined by
\begin{equation}
\begin{array}{ll}
Q_1^p=(\bar{b}_i p_i)_{V-A} (\bar{p}_j s_j)_{V-A},  &
Q_2^p=(\bar{b}_i p_j)_{V-A} (\bar{p}_j s_i)_{V-A},  \\
Q_3=(\bar{b}_i s_i)_{V-A} \sum_q (\bar{q}_j q_j)_{V-A},  &
Q_4=(\bar{b}_i s_j)_{V-A} \sum_q (\bar{q}_j q_i)_{V-A},  \\
Q_5=(\bar{b}_i s_i)_{V-A} \sum_q (\bar{q}_j q_j)_{V+A},  &
Q_6=(\bar{b}_i s_j)_{V-A} \sum_q (\bar{q}_j q_i)_{V+A},  \\
Q_7=(\bar{b}_i s_i)_{V-A} \sum_q\frac{3}{2}e_q (\bar{q}_j q_j)_{V+A},  &
Q_8=(\bar{b}_i s_j)_{V-A} \sum_q\frac{3}{2}e_q (\bar{q}_j q_i)_{V+A},  \\
Q_9=(\bar{b}_i s_i)_{V-A} \sum_q\frac{3}{2}e_q (\bar{q}_j q_j)_{V-A},  &
Q_{10}=(\bar{b}_i s_j)_{V-A} \sum_q\frac{3}{2}e_q (\bar{q}_j q_i)_{V-A},
\end{array}
\label{operator}
\end{equation}
where $i,j$ are color indices,
$e_q$ is the electric charge of the quark,
$(\bar{q}_1q_2)_{V\pm A}=\bar{q}_1\gamma_\mu(1\pm \gamma_5)q_2$ and
$q=u,d$.

The  topological amplitudes are written in terms of the Wilson coefficients
in the standard operator basis as~\cite{Buchalla:1995vs, Beneke:2001ev}
\bea
A(B^+ \to \pi^+ K^0) &=&
-\lambda_t\left[ \left(a_4-\frac{1}{2}a_{10}\right)
+ r^K_\chi\left(a_6-\frac{1}{2}a_{8}\right)
\right] A_{\pi K}
 , \nonumber\\
\sqrt{2} A(B^+ \to \pi^0 K^+) &=&
A({B}^0 \to \pi^- K^+)
-\left[ \lambda_u a_2 +\frac{3}{2}\lambda_t\left(a_7 -a_9\right)
\right] A_{K \pi  }
\nonumber\\
A({B}^0 \to \pi^- K^+) &=&
-\left[ \lambda_u a_1 -\lambda_t\left(a_4 +a_{10}\right)
- \lambda_t r^K_\chi\left(a_6+ a_{8}\right)
\right] A_{\pi K}
    , \nonumber\\
\sqrt{2} A({B}^0 \to \pi^0 {K}^0) &=&
A(B^+ \to \pi^+ K^0)+\sqrt{2} A(B^+ \to \pi^0 K^+)-A({B}^0 \to \pi^- K^+)
,
\label{eq:all2a}
\eea
where  $a_i = C_i + C_{i\pm 1}/3$ with $+(-)$ sign  for
odd (even) $i$.
We can specify each penguin contributions as \cite{Beneke:2001ev},
\bea
P'_{tc} &=& -\lambda_t \Big[a_4 + r^K_\chi a_6
 \Big]
A_{\pi K}, \label{eq:P:a}\\ \nonumber
P'_{\rm EW} &=&
\frac{3}{2} \lambda_t \Big[a_7 -  a_9
 \Big] A_{K \pi}, \\ \nonumber
P^{'C}_{\rm EW} &=& -\frac{3}{2} \lambda_t
\Big[a_{10} + r^K_\chi a_8
\Big]
A_{\pi K} ,
\\ \no
T' &=& \lm_u a_1 A_{\pi K},
\\ \no
C' &=& \lm_u a_2 A_{K \pi}.
\label{sm1}
\eea
where $r^K_\chi = 2 m_K^2/m_b(m_s + m_q)$,
$m_q = (m_u + m_d)/2$,
$A_{\pi K (K \pi)} = G_F (m_B^2 - m_{\pi(K)}^2)
F_0^{\pi (K)}
f_{K(\pi)}/\sqrt{2}$, and $F_0^{\pi (K)}
\simeq 0.3$ are semileptonic form factors for $B$ decays \cite{AlHaydari:2009zr}.

\section{Review of the Bulk Randal-Sundrum Model }
\label{sec:RS}

The RS model is based on
a 5D warped spacetime with the metric\;\cite{Randall:1999ee}
\begin{equation}
ds^2= \frac{1}{(kz)^2}(dt^2-dx^2 - dz^2),
\end{equation}
where the fifth dimension  $z$ is compactified between $1/k<z<1/T$. Here $k\simeq M_{\rm Pl}$ and
$T$ is the natural cut-off of the gauge theory
at TeV scale. Two boundaries $z_{UV}=1/k$ and $z_{IR}=1/T$ are called the Planck brane and the TeV brane, respectively.

For $SU(2)$ custodial symmetry,
we adopt the model suggested by  Agashe {\it et.al.} in Ref. \cite{Agashe:2003zs},
based on the gauge structure of
\gauge.
The custodial symmetry is guaranteed by the bulk $SU(2)_R$ gauge symmetry.
The bulk gauge
$SU(2)_R$ symmetry is broken into $U(1)_R$
by the orbifold boundary conditions on
the Planck brane:
charged $SU(2)_R$ gauge fields have $(-+)$ parity.
The $U(1)_R \times U(1)_{X} $ is
spontaneously broken into $U(1)_Y$  on the Planck brane
and  the Higgs field localized on the TeV brane is responsible to the breaking of
$SU(2)_L \times U(1)_Y$ to $U(1)_{\rm EM}$.

The action for a 5D
gauge fields is
\begin{eqnarray}
\label{eq:S-A}
S_{\mathrm{gauge}} = \int d^4 x dz
\sqrt{G}\left[-\frac{1}{4}g^{MP}g^{NQ}F^{a}_{MN}F^{a}_{PQ}\right],
\end{eqnarray}
where $G$ is the determinant of the AdS metric $g^{MN}$,
and $F^a_{MN}=\partial_M A^a_N- \partial_N A^a_M+g_5 f^{abc}A^b_MA^c_N$.
The 5D action for the gauge interactions of a bulk fermion $\hat\Psi(x,z) \equiv \Psi(x,z)/(k z)^2 $ is
\beq S_{\rm int} =
\int \di^4 x \, \di z \sqrt{G} \frac{g_{5i}}{\sqrt{2k}} \bar{\hat\Psi}(x,z)i
\Gamma^M A^a_M(x,z)T^a \hat\Psi(x,z),
\eeq
where $g_{5i}$ is the 5D
dimensionless gauge coupling ($g_{5s}, g_{5L}, g_{5R}, g_{5X}$) for each gauge group ($SU(3)_c, SU(2)_L, SU(2)_R, U(1)_X$)
and $\Gamma^M=(\gamma^\mu,i\gamma_5)$.

A bulk gauge field
$ A_{\nu} (x,z)$ and a bulk fermion field $\hat{\Psi}(x,z)$
are expanded in terms of KK modes by
%
\begin{eqnarray}
\label{eq:KKexpA} A_{\nu} (x,z)
&=& \sqrt{k} \sum_n A_\nu^{(n)}(x)
\fan (z),
\\ \no
\hat{\Psi}(x,z) &=&
\sqrt{k} \sum_{n} \left[ \psi_L(x) f_{L}(z) + \psi_R(x)
f_R(z) \right],
\end{eqnarray}
where the mode functions of
$\fan (z)$ and $f_{L}^{(0)}(z,c) =f_{R}^{(0)}(z,-c)$
are referred to
Ref. \cite{Chang:2005ya}.
Here $c$ is defined through the 5D Dirac mass $m_D =
{\rm sign} (y) ck$ and $z = e^{k |y|}$.
Note that a massless SM fermion corresponds to the zero mode with $(++)$ parity.
Since $\Psi_L$ has always opposite parity of $\Psi_R$,
the left-handed SM fermion is the zero mode of a 5D
fermion whose left-handed part has $(++)$ parity (the
corresponding
right-handed part has automatically $(--)$ parity which cannot have a zero mode).

Due to the gauge structure of \gauge,
the right-handed SM fermions belong to a $SU(2)_R$ doublet, and couple to
$SU(2)_R$ gauge bosons with $(-+)$ parity.
As a result, the whole quark sector is
\begin{equation} Q_i =\left(
       \begin{array}{c}
         u_{iL}^{(++)} \\
         d_{iL}^{(++)} \\
       \end{array}
     \right),
     \quad
U_i =\left(
       \begin{array}{c}
         u_{iR}^{(++)} \\
         D_{iR}^{(-+)} \\
       \end{array}
     \right),
     \quad
D_i =\left(
      \begin{array}{c}
        U_{iR}^{(-+)} \\
        d_{iR}^{(++)} \\
      \end{array}
    \right),
\end{equation}
where $i$ is the generation index.
Dirac mass parameters ($c_{Q_i}$, $c_{U_i}$, $c_{D_i}$)
determine their
mode functions, KK mass spectra, and coupling strength with KK gauge bosons.

On the TeV brane, the SM fermion
mass is generated as the localized Higgs field develops its VEV
of $\langle H \rangle = v\simeq 174\gev$.
The SM mass matrix for a fermion $f (= u,d,\nu,e)$ is
\begin{equation}
\label{eq:Mf:general}
\big(M_f \big)_{ij}= v \lambda^f_{5ij}
\left. \frac{k}{T} \, f_R^{(0)}(z,c_{Ri}) f_L^{(0)}(z,c_{Lj})
\right|_{z=1/T}
\equiv v \lambda^f_{5ij} F_R(c_{f_{Ri}}) F_L(c_{f_{Lj}}),
\end{equation}
where $i,j$ are the generation indices,
$\lambda^f_{5ij}$ are the 5D (dimensionless) Yukawa couplings,
and $F_L(c)=F_R(-c)$ is defined by
\beq
F_{L}(c) \equiv \frac{ f_{L}^{(0)}\left( {1}/{T},c \right) }{\es^{1/2}}
,
\eeq
where $\es = T/k$.
The mass eigenstates of the SM fermions
involve two independent mixing matrices, defined by
\begin{equation}
\label{eq:chi-psi-mix}
\chi_{fL}=U^\dagger_{fL} \psi^{(0)}_{fL},\quad \chi_{fR}=U^\dagger_{fR} \psi^{(0)}_{fR}.
\end{equation}
Note that the observed mixing matrix is a multiplication of two
independent mixing matrices
such that $V_{\rm CKM}  = U_{uL}^\dagger U_{dL}$
and $\pmns = U_{e L}^\dagger U_{\nu L}$.

If bulk Dirac mass parameters
and 5D Yukawa couplings are given \textit{a priori},
we could predict all the mass spectra
and mixing matrices as well as the
couplings with KK gauge bosons.
Without those crucial knowledge,
we have to take
the opposite way, {\it i.e.},
deducing them from the observation.
The problem is that the number of observations is
not enough to fix all the model parameters.
In the previous work,
we have developed a theory
based on the following two natural \textsl{assumptions}:
\begin{enumerate}
    \item For all fermions, 5D Yukawa couplings have
the same order of magnitude $\lm_5^f \sim \mathcal{O}(1)$.
    \item When explaining the observed
mixing matrix $V_{\rm CKM}  = U_{uL}^\dagger U_{dL}$
and $\pmns = U_{e L}^\dagger U_{\nu L}$,
no order-changing by cancellation is allowed.
\end{enumerate}
The \textsl{assumption}-1 yields
anarchy type fermion mass matrix,
which naturally explains the large top quark mass
$v \simeq 174\gev$.
Other small SM fermion masses
are generated by controlling $c$'s.
The \textsl{assumption}-2 is consistent with the spirit of no fine-tuning.

In Ref. \cite{Chang:2005ya},
we have shown that the above two natural assumptions
determine the nine bulk mass parameters
within a well-defined regions as
\begin{eqnarray}
\label{eq:c:quark} c_{Q_1} &\simeq &0.61,\quad
\, c_{Q_2}  \simeq 0.56 ,\quad c_{Q_3}
\simeq 0.3^{+0.02}_{-0.04},
\\ \no
c_{U_1} &\simeq& -0.71 ,\ c_{U_2} \simeq -0.53 ,\ 0\lsim c_{U_3} \lsim
0.2,
\\ \no
c_{D_1} &\simeq& -0.66 ,\ c_{D_2} \simeq -0.61 ,\ c_{D_3} \simeq -0.56
..
\end{eqnarray}
Recently phenomenological constraint on the value of $c_{Q_3}$
has been studied, focused on the anomalous coupling of
$Z b \bar{b}$ vertex~\cite{Casagrande:2010si}:
$c_{Q_3}$ cannot be smaller than 0.3.
Combined with our constraint based on the two naturalness
assumptions,
we consider the case of $c_{Q_3} = 0.3 \sim 0.32$
in what follows.

The
quark mixing matrices are
\begin{eqnarray}
\label{eq:UqL:UqR}
 \left( U_{qL} \right)_{ij(i\leq j)} \approx
\frac{F_L(c_{Q_i})}{F_L(c_{Q_j})},\quad \left( U_{qR}
\right)_{ij(i\leq j)} \approx \frac{F_R(c_{A_i})}{F_R(c_{A_j})},
\quad A=U,D.
\label{eq:ULUR:FLFR}
\end{eqnarray}
Then our mixing matrices show the following
order of magnitude behaviors:
\bea
\label{eq:U:lambda}
U_{uL} &\simeq& U_{dL} \simeq
\left(
  \begin{array}{ccc}
    1 & \lm & \lm^3 \\
    \lm & 1 & \lm^2 \\
    \lm^3 & \lm^2 & 1 \\
  \end{array}
\right) + \mathcal{O}\left( \lm^4 \right),
\\ \label{eq:UuR}
U_{uR} &\simeq&
\left(
  \begin{array}{ccc}
    1 & 0 & 0 \\
    0 & 1 & \lm^2 \\
    0 & \lm^2 & 1 \\
  \end{array}
\right) + \mathcal{O}\left( \lm^4 \right),
\quad
U_{dR} \simeq
\left(
  \begin{array}{ccc}
    1 & \lm & \lm^2 \\
    \lm & 1 & \lm \\
    \lm^2 & \lm & 1 \\
  \end{array}
\right) + \mathcal{O}\left( \lm^3 \right).
\eea

As shall be shown below,
only $U_{uL}$ and $U_{dL}$ make dominant contributions
to $B\to K\pi$ decays.
Because of
high similarity of $U_{uL}$ and $U_{dL}$ to the CKM matrix,
we take the following assumptions of
\bea
\left( U_{qL} \right)_{ij}
= \kp_{ij} \left(V_{\rm CKM}\right)_{ij}.
\eea
In order to satisfy our naturalness assumptions, we require
\bea
\label{eq:kappa}
\frac{1}{\sqrt{2}} < \left| \kp_{ij}\right| < \sqrt{2}.
\eea

\section{Bulk RS model effects on $B \to K\pi$ decays}
\label{sec:effect}

In the bulk RS model, the mass eigenstate of a SM fermion is a mixture of
gauge eigenstates as in Eq.~(\ref{eq:chi-psi-mix}) and
we have FCNC mediated by KK gauge bosons.
In terms of gauge eigenstates, the four-dimensional (4D) gauge interactions with KK gauge modes $A^{a(n)}_\mu$
are
\beq
\mathcal{L}_{4D} \supset
g_{4j} \sum_{n=1}^\infty
\left(\hat{g}^\n_{L}(c_{i})\, \bar{\psi}_{iL}^{(0)} T^a\gamma^\mu
\psi_{iL}^{(0)}+\hat{g}^\n_{R}(c_{i})\, \bar{\psi}_{iR}^{(0)} T^a\gamma^\mu
\psi_{iR}^{(0)}\right) A^{a(n)}_\mu,
\eeq
where $g_{4j}=g_{5j}/\sqrt{kL}$ for $j=s,L,R,X$.
Since the bulk RS effects are suppressed by the
forth power of the KK mass,
we consider only the contribution of
the first KK mode of gauge bosons.
In what follows, therefore, we omit the KK mode number notation $(n)$.
Then the effective gauge couplings with the first KK gauge boson
are
\begin{eqnarray}
\label{eq:ghat:def}
\hat{g}_{L}(c_{f_i}) =
\hat{g}_{R}(-c_{f_i}) =\sqrt{k L} \int \di z k \left[
f_L^{(0)}(z,c_{f_i}) \right]^2 f^{(1)}_A(z)\equiv\hat{g}(c_{f_i})
..
\end{eqnarray}
Note that the effective coupling $\hat{g}(c)$ vanishes
if $c=1/2$.

The values of the bulk mass parameters $c$'s
in Eq.~(\ref{eq:c:quark}) fix the $\hat{g}$ values as
\bea
\label{eq:ghat:value}
&&
\hat{g} (c_{Q_1}) = -0.192,
\quad
\hat{g}(c_{Q_2}) = -0.179,
\quad
\hat{g}(c_{Q_3}) = 1.797 \sim 1.974,
\\ \no
&&
\hat{g} (c_{U_1}) = -0.193,
\quad
\,\hat{g}(c_{U_2}) = -0.133,
\quad
\hat{g}(c_{U_3}) \,= 2.759\sim3.948,
\\ \no
&&
\hat{g} (c_{D_1}) = -0.193,
\quad
\hat{g}(c_{D_2}) = -0.192,
\quad
\hat{g}(c_{D_3}) = -0.179,
\eea
where, for example, $\hat{g}=1.974$ for $c_{Q_3} =0.3$
and $\hat{g}=1.797$ for $c_{Q_3} =0.32$.
It can be seen that $\hat{g}(c_{Q_3})$ and $\hat{g}(c_{U_3})$
are dominant over all the other $\hat{g}$'s
of the order of $\lm$.

The relevant FCNC processes
mediated by the
first KK gauge bosons are described by the following Lagrangian:
\begin{eqnarray}
\no
\mathcal{L}_{4D} \!\!\!&=&\!\!\!
-g_s  \left(  K^{uL}_{lm}\bar{u}_{lL}T^a  \gamma^\mu u_{mL}
+K^{dL}_{lm}\bar{d}_{lL}T^a  \gamma^\mu d_{mL}
+ K^U_{lm}\,\bar{u}_{lR}T^a  \gamma^\mu u_{mR}
+ K^D_{lm}\,\bar{d}_{lR}T^a  \gamma^\mu d_{mR}
\right) G^{a(1)}_\mu
\\ \nonumber
\!\!\!&&\!\!\!
-\frac{1}{2} \biggl[
g\left(K^{uL}_{lm}\,\bar{u}_{lL}  \gamma^\mu u_{mL}
-K^{dL}_{lm}\,\bar{d}_{lL}  \gamma^\mu d_{mL}
\right) W^{(1)}_{3L\mu}
\\ \!\!\!&&~~
+
\tilde{g}
\left(K^{U}_{lm}\,\bar{u}_{lL}  \gamma^\mu u_{mL}
+K^{D}_{lm}\,\bar{d}_{lR}  \gamma^\mu d_{mR}
\right) W^{(1)}_{3R\mu} \nonumber
\\ \label{eq:eff:Lg}
\!\!\!&&~~
+g_X
\left(K^{uL}_{lm}\,\bar{u}_{lL}  \gamma^\mu u_{mL}
+ K^{dL}_{lm}\,\bar{d}_{lL}  \gamma^\mu d_{mL}
+K^{U}_{lm}\,\bar{u}_{lL}  \gamma^\mu u_{mL}+
 K^D_{lm}\,\bar{d}_{lR}  \gamma^\mu d_{mR}
\right) B^{(1)}_{X\mu} \biggl],
\end{eqnarray}
where the subscript $l,m$ are the generation indices
($l,m=1,2,3$), and the 4D gauge couplings are
$g_4 = {g_{5L}}/{\sqrt{kL}}$.
The effective mixing matrices $K$'s are
\begin{eqnarray} \label{eq:Kdef}
K^{qL}_{lm}&=& \sum_{k=1}^3 \big( U_{qL}^\dagger \big)_{lk}
\,\hat{g}(c_{Q_k}) \left( U_{qL} \right)_{km} ,\quad
\hbox{ for } q=u,d,
\nonumber\\
K^U_{lm}&=& \sum_{k=1}^3 \big( U_{uR}^\dagger \big)_{lk}
\,\hat{g}(-c_{U_k}) \left( U_{uR} \right)_{km} \,,
\nonumber\\
K^D_{lm}&=& \sum_{k=1}^3 \big( U_{dR}^\dagger \big)_{lk}
\,\hat{g}(-c_{D_k}) \left( U_{dR} \right)_{km} \,.
\end{eqnarray}

We first estimate the value of the elements of
$K$'s.
Since $\hat{g}(c_{Q_1}) \approx \hat{g}(c_{Q_2}) \ll \hat{g}(c_{Q_3})$
as in Eq.~(\ref{eq:ghat:value}),
the dominant elements of $K_{qL}$ are
\bea
\label{eq:K23}
K^{qL}_{32}&\simeq& \big( U_{dL} \big)_{33}\big( U_{dL} \big)_{32} \hat{g}(c_{Q_3})
\sim \lm^2 \hat{g}(c_{Q_3}),
 \nonumber\\
K^{qL}_{31}&\sim&
 \lm^3 \hat{g}(c_{Q_3}),
\quad
K^{qL}_{11}\sim\hat{g}(c_{Q_1}),
\quad
K^{qL}_{12} \sim \lm\hat{g}(c_{Q_1}).
\eea
In addition, the condition of
$\hat{g}(c_{D_1}) \approx \hat{g}(c_{D_2}) \approx \hat{g}(c_{D_3})$, up to $\mathcal{O}(\lm^2)$,
leads to diagonal $K^D$ up to $\mathcal{O}(\lm^4)$:
\bea
K^D_{ij(i\neq j)} = 0 .
\eea
Note that these vanishing off-diagonal elements of the right-handed
quarks
ensure the validity of the operator expansions
in the effective Hamiltonian Eq.~(\ref{eq:H:def:C}).
The diagonal elements are
\bea
\label{eq:K11}
K^D_{11} \sim \hat{g}(c_{D_1}),
\quad
K^U_{11}\sim \hat{g}(c_{U_1}).
\eea

Finally we have the
effective Hamiltonian for $B \to K \pi$ decay, given by
\begin{eqnarray}
\no
 {\cal H}_{\rm RS}
&\simeq&
\frac{g_s^2 K^{dL}_{32}}{8 M_{KK}^2} \biggl[
 (\bar{b}_i s_j)_{_{V-A}}
\left\{
K^{uL}_{11} (\bar{u}_j u_i)_{_{V-A}}
+
K^{dL}_{11} (\bar{d}_j d_i)_{_{V-A}}
+
K^{U}_{11} (\bar{u}_j u_i)_{_{V+A}}
+
K^{D}_{11} (\bar{d}_j d_i)_{_{V+A}}
\right\}
\\ \no && ~~~~~
- \frac{1}{3}
 (\bar{b} s)_{_{V-A}}
\left\{
K^{uL}_{11} (\bar{u} u)_{_{V-A}}
+
K^{dL}_{11} (\bar{d} d)_{_{V-A}}
+
K^{U}_{11} (\bar{u} u)_{_{V+A}}
+
K^{D}_{11} (\bar{d} d)_{_{V+A}}
\right\}
\biggl]
\\ \label{eq:eff:H} && \!\!\!\!
-\frac{g^2 K^{dL}_{32}}{16 M_{KK}^2}
 (\bar{b} s)_{_{V-A}}
\left\{
K^{uL}_{11} (\bar{u} u)_{_{V-A}}
-
K^{dL}_{11} (\bar{d} d)_{_{V-A}}
 \right\}
\\
&&\!\!\!\!
-\frac{g_X^2K^{dL}_{32}}{16  M_{KK}^2}
 (\bar{b} s)_{_{V-A}}
\left\{
K^{uL}_{11} (\bar{u} u)_{_{V-A}}
+
K^{dL}_{11} (\bar{d} d)_{_{V-A}}
+
K^{U}_{11} (\bar{u} u)_{_{V+A}}
+
K^{D}_{11} (\bar{d} d)_{_{V+A}} \right\}
\,,\nonumber
\label{effective2}
\end{eqnarray}
where $i,j$ are the color indices
and $M_{KK}$ is the first KK gauge boson mass.
Here we have included only
the most dominant terms proportional to $K_{32}^{dL}$
since the values of $\hat{g}(c_{Q_3})$ and $\hat{g}(c_{U_3})$
are
much larger than those of the other $\hat{g}$'s.

Matching the NP contribution to Wilson coefficients from
Eq.~(\ref{effective}) and Eq.~(\ref{operator}), we get even
$C_i$'s of
\bea
C_4&=&  -\frac{B_{RS}}{\lmt} g_s^2
\left(K^{uL}_{11} +2 K^{dL}_{11}
\right),
\quad
C_6=  -\frac{B_{RS}}{\lmt} g_s^2
\left(K^{U}_{11} +2 K^{D}_{11}
\right),
\\ \no
C_8&=&  -2\frac{B_{RS}}{\lmt} g_s^2
\left(K^{U}_{11} - K^{D}_{11}
\right),
\quad
C_{10} =  -2\frac{B_{RS}}{\lmt} g_s^2
\left(K^{uL}_{11} - K^{dL}_{11}
\right),
\eea
and odd $C_i$'s of
\bea
C_3 &=& \frac{B_{RS}}{\lmt} \left\{
\left(\frac{g_s^2}{3}+ \frac{g^2+g_X^2}{2}\right)K_{11}^{uL}
+ 2\left(\frac{g_s^2}{3}+\frac{g_X^2-g^2}{2} \right)K_{11}^{dL}
\right\},
\\ \no
C_5 &=& \frac{B_{RS}}{\lmt}
\left( \frac{g_s^2}{3}  + \frac{g_X^2}{2}
\right)
\left(K^{U}_{11} + 2 K^{D}_{11}
\right),
\\ \no
C_7 &=& \frac{B_{RS}}{\lmt}
\left( \frac{2}{3}g_s^2 + g_X^2 \right)
\left(K^{U}_{11} - K^{D}_{11}
\right),
\\ \no
C_9 &=& \frac{B_{RS}}{\lmt} \left\{
\left(\frac{2g_s^2}{3}+ g^2+g_X^2\right)K_{11}^{uL}
- \left(\frac{2 g_s^2}{3}-g^2+g_X^2 \right)K_{11}^{dL}
\right\}.
\eea
Here $B_{RS}$ denotes the common NP factor, given by
\beq
\label{eq:BRS}
B_{RS} = \frac{\sqrt{2} K_{32}^{dL} }
{24 G_F M_{KK}^2}=\frac{1}{3 g^2}
\left( \frac{m_W}{M_{KK}} \right)^2
K_{32}^{dL}.
\eeq

The leading NP contributions in the same form as the penguin diagrams
of Eq.~(\ref{eq:P:a}) become
\bea
\label{eq:P:NP}
P'_{NP} &=& B_{RS} \left[
\left( \frac{8}{9}g_s^2 - \frac{g_X^2}{6}\right)
\left\{ K_{11}^{uL} + 2 K^{dL}_{11}
+ r^K_\chi
\left(K_{11}^{U} + 2 K^{D}_{11} \right)
\right\}
\right.
\\ \no && ~~~~~
\left.
-\frac{g^2}{6}\left(K_{11}^{uL} -2 K^{dL}_{11}\right)
\right]
A_{\pi K} ,\\
\label{eq:Pew:NP}
P'_{EW,NP} &=& -\frac{3}{2}B_{RS}
\left[
g_X^2
\left(
K^{uL}_{11} - K^{dL}_{11}-K^U_{11}+K^D_{11}
\right)
+ g^2 (K_{11}^{uL}+K_{11}^{dL})
\right]
A_{K\pi },
\\ \label{PC:NP2}
P^{'C}_{\rm EW,NP} &=&
\frac{3}{2}B_{RS} \left[
\left( \frac{16}{9}g_s^2 - \frac{g_X^2}{3}\right)
\left\{ K_{11}^{uL} - K^{dL}_{11}
+ r^K_\chi
\left(K_{11}^{U} - K^{D}_{11} \right)
\right\}
\right.
\\ \no && ~~~~~
\left.
-\frac{g^2}{3}\left(K_{11}^{uL}+K^{dL}_{11}\right)
\right]
A_{\pi K}
 .
\eea
The SM coupling of U(1)$_Y$ is $g_Y=g_X\tilde g/\sqrt{g_X^2+\tilde g^2}$.

If we redefine
\bea
\bar P \equiv P'+P'_{NP}-\frac{1}{3}( P^{'C}_{\rm EW}+P^{'C}_{\rm EW,NP})  ,
\eea
Eq.~(\ref{eq:all2}), ignoring ${\cal O}(\lambda^2)$ terms, becomes
\bea
A(B^+ \to \pi^+ K^0) &=&\bar P
 , \nonumber\\
\sqrt{2} A(B^+ \to \pi^0 K^+) &=& -\bar P - T' -  P'_{\rm EW}
-P^{'}_{\rm EW,NP}-P^{'C}_{\rm EW,NP}  ,
\nonumber\\
A({B}^0 \to \pi^- K^+) &=& -\bar P  - T'-P^{'C}_{\rm EW,NP}
    , \nonumber\\
\sqrt{2} A({B}^0 \to \pi^0 {K}^0) &=&
\bar P - P'_{\rm EW} -P^{'}_{\rm EW,NP}  .
\label{eq:all4}
\eea
Considering the fact that the NP contribution should not affect the well-measured
BR's of $B \to K \pi$ decays,
it is reasonable to assume that
all NP contributions should be smaller than $P'_{tc}$.
In the case where the strong CP phases of $P'_{NP}$ and
$P^{'C}_{\rm EW,NP}$ are the same,
the ratio of  the NP contribution of $P'$ to the SM  $P'_{tc}$ in this model is
\bea
\frac{P'_{NP}-P^{'C}_{\rm EW,NP}/3}{P'_{tc}}
&\sim& - \frac{B_{RS} (8g_s^2/3-g_X^2/2)
\left\{ K_{11}^{dL}
+ r^K_\chi  K^{D}_{11}
\right\} }{\lm_t (a_4 + r_X a_6)}
\\ \no
&\sim& - \lm  C \left(
\frac{ 300 \gev}{M_{KK}}
\right)^2,
\eea
where  $K_{32}^{dL}  \sim K^{D}_{11}
\sim \lm,
K_{32}^{dL} \sim \lm_t\hat g(c_{Q3})  $ and $|a_4+r_X a_6|\sim 0.05$ to $0.1$  \cite{Beneke:2001ev,Li:2005kt}.
The constant $C$ is ${\cal O}(1)$ and less than 10
in conservative estimation.
The KK mass over 1 TeV suppresses
new contribution to the branching ratios of $B \to K \pi$ decays.

On the other hand,
the $\Delta A_{CP}$, defined in Eq.~(\ref{eq:dACP}),
can be explained by new
contribution to the $P'_{\rm EW}$.

We adopt the simplifying notation of
\bea
\bar P =- |P'_{tc}| \eta_{NP},
\eea
where $\eta_{NP}\simeq (1+ P'_{NP}/|P'_{tc}| -P^{'C}_{\rm EW,NP}/3|P'_{tc}|)$.
The contribution of
$\eta_{NP}$ to $A_{\rm CP}$ is  suppressed, if $M_{KK}>1$ TeV,
 $P'_{NP}/|P'_{tc}| \ll \lm$, and $ P^{'C}_{\rm EW,NP}/3|P'_{tc}| \ll \lm$.
The NP effect can be written as
\bea
A(B^+ \to \pi^+ K^0) &=& -|P'_{tc}| \eta_{NP}
 , \nonumber\\
\sqrt{2} A(B^+ \to \pi^0 K^+) &=&
|P'_{tc}|\eta_{NP}(1-r_{EW}e^{i\delta_{EW}}-r_Te^{i\gamma}e^{i\delta_T}-r_{1}e^{i\phi_1}e^{i\delta_1}-r_2e^{i\phi_2}e^{i\delta_2})  ,
\nonumber\\
A({B}^0 \to \pi^- K^+) &=&  |P'_{tc}|\eta_{NP}(1-r_Te^{i\gamma}e^{i\delta_T}-r_2e^{i\phi_2}e^{i\delta_2})   , \nonumber\\
\sqrt{2} A({B}^0 \to \pi^0 {K}^0) &=& -|P'_{tc}|\eta_{NP}(1+r_{EW}e^{i\delta_{EW}}+r_1e^{i\phi_1}e^{i\delta_1}) ,
\label{eq:all3}
\eea
where  $r_{EW} \equiv |P'_{EW}/P'_{tc}|$, $r_T\equiv |T'/P'_{tc}|$, $r_1 \equiv |P'_{\rm EW,NP}/P'_{tc}|$, $r_2\equiv |P^{'C}_{\rm EW,NP}/P'_{tc}|$,
$\phi_i$'s are the weak phases and $\delta_i$'s are
strong phases.
The difference between two $A_{\rm CP}$
is
\bea
\Delta A_{\rm CP}
&\simeq& 2 r_1\sin\delta_1 \sin\phi_1,
\label{r1}
\eea
where $r_1$ is
\bea
\label{eq:r1}
r_1 &\simeq&
\frac{1}{2 g^2 }\left|(K_{11}^{uL} - K_{11}^{dL}-  K_{11}^{U}+ K_{11}^D)g_X^2+(K_{11}^{uL} + K_{11}^{dL})g^2 \right|
\frac{\hat g(c_{Q3}) |\kappa_{33}^*\kappa_{32}|}{|a_4
+ r^K_\chi a_6|}\frac{f_\pi}{f_K}
\frac{m_W^2}{M_{KK}^2}
\\ \no
&\lsim & \frac{1}{2g^2 }\left[(\lambda g_X)^2+g^2\right] \frac{m_W^2}{M_{KK}^2}\,R_1 ,
\eea
where $R_1 = 2  \hat{g}(c_{Q1})
\hat g(c_{Q3}) |\kappa_{33}^*\kappa_{32}|f_\pi
/(f_K|a_4+ r^K_\chi a_6|)$.
Here $\kappa_{ij}$ is defined in Eq.~(\ref{eq:kappa}) and $\lambda\simeq 0.22$.
The value of 
$(K_{11}^{uL} - K_{11}^{dL}-  K_{11}^{U}+ K_{11}^D)$ has a suppressed value of
$ \kappa^2\lambda^2 \hat{g}(c_{Q1})$ 
because of the semi-diagonal mixing matrices of
$U_{qL,qR}$ as in Eq.~(\ref{eq:U:lambda}),
while
$K_{11}^{uL} + K_{11}^{dL}\sim 2 \hat{g}(c_{Q1})$.  
We have used the mixing parameters
in Eqs.(\ref{eq:U:lambda}-\ref{eq:kappa}).
In our choices of input parameters, the range of $R_1$ is from 4 to 20, depending on the effective mixing matrices $K$'s,
$|\kappa_{33}^*\kappa_{32}|$ and the SM
values of $a_4$ and $ a_6$ which are around $0.05\sim 0.07$.

\begin{figure}[t!]
\centering
  \includegraphics[scale=1]{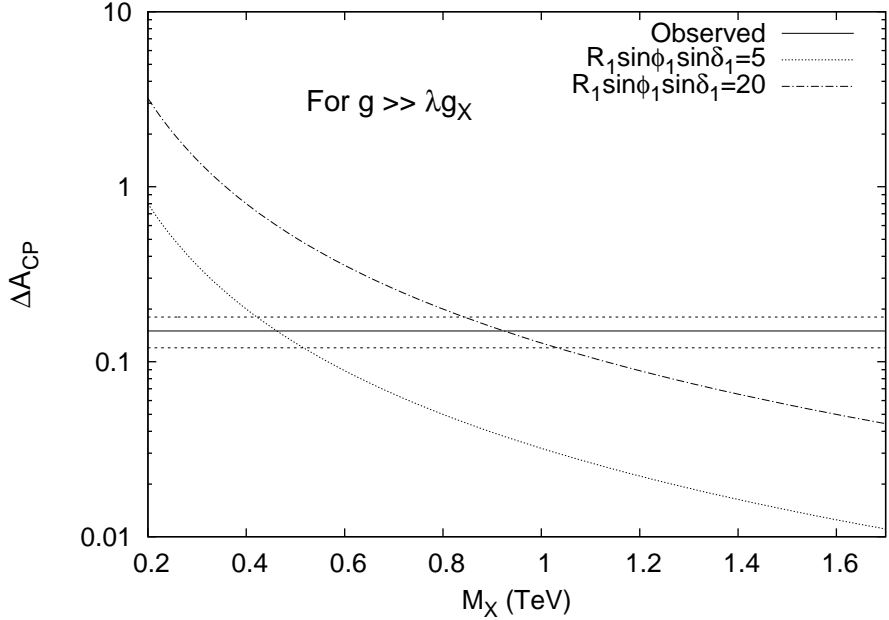}
\caption{\label{fig:DACP1}\small
$\Delta A_{CP}$ for $R_1 \sin \dt_1 \sin \phi_1 =5$
and $R_1 \sin \dt_1 \sin \phi_1 =20$ when $g\gg \lambda g_X$.
The horizontal lines are the observed $\Delta A_{CP}$
with the $ 1 \sigma$ experimental error.
}
\end{figure}

Brief review about the phenomenological constraints
on $M_{KK}$ is in order here.
In the minimal custodial RS model with an TeV-brane localized
Higgs field and anarchic 5D Yukawa couplings,
the constraints from $\epsilon_K$, $b\to s \gamma$
and $B^0 \bar{B}^0$ mixing put the lower bound of KK mass gauge boson above
 $10 \tev$~\cite{Csaki,Buras,UTfit,Chang:2006si}.
Later, it has been suggested that $M_{KK}$ can be lowered down to $\sim$5 TeV if
we release the Higgs field in the  bulk
and include the one-loop matching of gauge
couplings~\cite{Agashe, Gedalia}.
However, if we allow mild tuning in the 5D Yukawa couplings,
both models with the Higgs filed on the TeV-brane  or
in the bulk accommodate sizable region in parameter space
for $M_{KK}>3\tev$ \cite{Buras}.
In our model which allows mild tuning in terms of $\kp$,
we accept the phenomenological bound of
$M_{KK}>3\tev$, which is in marginal reach of the LHC.

\begin{figure}[t!]
\centering
\includegraphics[scale=1]{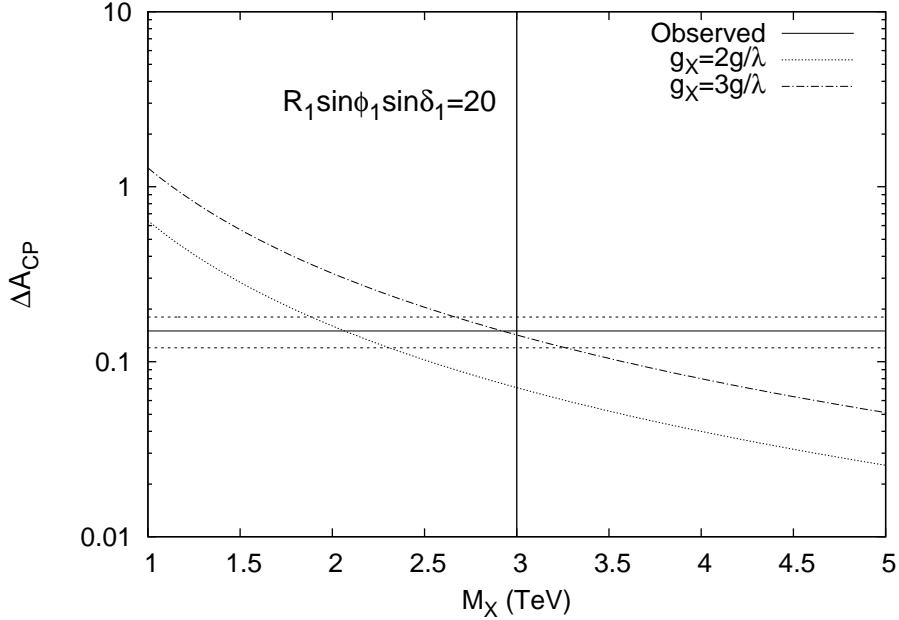}
\caption{\label{fig:DACP2}\small
The maximal $\Delta A_{CP}$ for $R_1 \sin \dt_1 \sin \phi_1 =20$
and $g_X=2 g/\lambda $ and $g_X = 3 g/\lambda$.
The horizontal lines are the observed $\Delta A_{CP}$
with the $ 1 \sigma$ experimental error.
}
\end{figure}

In Fig.\ref{fig:DACP1},
we plot $\Delta A_{\rm CP}$ for $R_1 \sin\delta_1 \sin\phi_1=5,20$ with $\lambda g_X$ ignored.
The horizontal lines are the observed $\Delta A_{CP}$
with the $ 1 \sigma$ experimental error.
As expected, $\Delta A_{\rm CP}$ decreases with
increasing $M_{KK}$.
For small $R_1\sin\delta_1 \sin\phi_1=5$,
the new contribution to $\Delta A_{\rm CP}$ is too small
for $M_{KK}>1\tev$. 
In order to explain $\Delta A_{\rm CP}$,
the scale $M_{KK}$ should be around $1\tev$ even for
the maximal value of $R_1$ and CP phases.
This low $M_{KK}$ is excluded 
by other phenomenological constraints.

In Fig.\ref{fig:DACP2}, we explore the possibility to explain 
the observed $\Delta A_{\rm CP}$ 
with $M_{KK} \simeq 3\tev$ 
in rather extreme case of large $g_X$. 
We found that this is feasible 
only when $U(1)_X$ coupling is large enough such as
$g_X \simeq 3g/\lambda$ and the 
contribution from $g_X$ sector in Eq.(\ref{eq:r1}) is maximal.
This large value of $g_X$ is marginally allowed by 
the pertubativity.

Large $g_X$ gives additional and possibly significant contributions to other experiments.
We notice that
four quark vertices such as $B^0 - \bar{B}^0$ mixing 
and $B \to  K \phi$ do not have severe contributions unless $g_X$ is not too larger than $g_s$,
because there are 
8 gluon modes.
We also consider the most stringent lepton number violating processes such as
$\mu \to 3 e$ and
$B \to K l^+l^-$ decays
where the dominant RS contribution
to decay amplitudes
is proportional to $\{g^2-(3g_X)^2\} \hat{g}(c_{Li})$ \cite{Chang:2007uz}.
These NP contributions are  proportional to
$\hat{g}(c_{Li})$ where $c_{Li}$'s ($i=e,\mu,\tau$)
are the bulk Dirac mass
parameters for the lepton doublets.
Since $\hat{g}(c)$ defined in Eq.~(\ref{eq:ghat:def})
vanishes if $c=1/2$,
it is required that $c_{Li}$ are extremely close to $1/2$  to suppress the large RS contribution.
Also, considering the perturbative nature of $U(1)_{B-L}$ gauge
couplings, it may be unnatural to assume  $g_X/g \gg 3$.

\section{Conclusions}
\label{sec:conclusions}
More precise data of $B \to K \pi$ decays have recently
raised
many interests.
While the branching ratios are well measured
and can be explained within the SM,
the direct CP asymmetries
in the $B^+ \to
\pi^0 K^+$ and $B^0 \to \pi^- K^+$
show significant deviation from the SM predictions.
This is called the
$\Dt A_{CP}$ puzzle in the $B \to K \pi$ decays.
We study this discrepancy in the
5D custodial bulk Randall-Sundrum
model where all the SM fields are in the bulk:
one exception is the localized Higgs boson field.

In this model, the Yukawa interactions with the localized
Higgs fields lead to non-zero masses for the SM fermions
which are the zeroth modes of the bulk fermion.
While the bulk gauge interactions are flavor-diagonal,
the Yukawa interactions generally mix the SM fermions
of different generations.
At tree level, the KK gauge bosons can mediate
FCNC.
We study these FCNC effects on four $B\to K \pi$ decays.
A custodial bulk RS model with two naturalness assumptions
has been adopted, where all the 5D Yukawa couplings
and the bulk Dirac mass parameters for the SM fermions
can be determined.

We showed that the operator expansions in the effective
Hamiltonian for $B \to K \pi$ take the same form to leading order
since the mixing matrix of the right-handed fermion fields
is almost diagonal.
To leading order, this model
has new contributions to the gluonic penguin amplitude $P'$,
color-favored and color-suppressed electroweak
penguin amplitudes $P_{EW}'$ and $P_{EW}^{\prime,C}$.
The well measured branching ratios of four $B \to K \pi$ decays
remain almost intact
if the KK mass scale is above 1 TeV.
On the contrary, the new contribution to
the color-favored electroweak penguin amplitude
$P_{EW,NP}'$ get NP contributions from the gluon KK mode exchange.
However its magnitude in a large portion of the model
parameter space is too small:
the discrepancy of $\Dt A_{CP}$
between the experimental data and the SM prediction
can be explained 
with $M_{KK} \simeq 1 \tev$, which contradicts
with other phenomenological constraints.
In rather extreme case of very large $g_X$, however, the
observed $\Dt A_{CP}$
can be explained with $M_{KK} \simeq 3 \tev$.

\acknowledgments
\noindent The work of S.C. was supported
by the Korea Research Foundation Grant. (KRF-2008-359-C00011).
The work of C.S.K.  was supported in part by Basic Science
Research Program through the NRF of Korea funded by MOEST
(2009-0088395), in part by KOSEF through the Joint Research
Program (F01-2009-000-10031-0).
The work of J.S. was supported by
the
WCU program through the KOSEF funded by the MEST
(R31-2008-000-10057-0).

\end{document}